# Transfer Learning Meets Embedded Correlated Wavefunction Theory for Chemically Accurate Molecular Simulations: Application to Calcium Carbonate Ion-Pairing

Xuezhi Bian[1] and Emily A. Carter[2,3]

[1]Department of Chemistry, Princeton University, Princeton, New Jersey 08544, United States

[2]Department of Mechanical and Aerospace Engineering, Princeton University, Princeton, New Jersey 08544, United States

[3]Andlinger Center for Energy and the Environment and Program in Applied and Computational Mathematics, Princeton University, Princeton, New Jersey 08544, United States

## Abstract

Achieving chemical accuracy for molecular simulations remains a central challenge in computational chemistry. Here, we present an embedded correlated wavefunction transfer learning (ECW-TL) framework for accurately simulating molecular dynamics in the condensed phase. ECW-TL incorporates high-level electron exchange and correlation effects in ECW theory while preserving training and computational efficiency of machine learned interatomic potentials. We demonstrate the framework on $Ca^{2+}$-$CO_3^{2-}$ ion pairing in aqueous solution, a key process underlying $CO_2$ mineralization in seawater. As proof of principle, we first show that finetuning a DFT-revPBE-D3(BJ) baseline model with embedded-DFT-SCAN data reproduces the DFT-SCAN free-energy surface within 1 kcal/mol across all solvation states. Extending the framework to embedded MP2 and localized natural-orbital CCSD(T) further refines the free-energy profile, revealing the crucial role of exact electron exchange and correlation in determining ion-pair stability and structure. ECW-TL thus provides a general, data-efficient route for transferring CW accuracy to large-scale simulations of complex aqueous and interfacial chemical processes.

# Introduction

*Ab initio* molecular dynamics (AIMD) simulation has become a fundamental tool for gaining understanding of microscopic phenomena in molecular and condensed-phase systems,[1,2] providing insights into many processes, including those of interest here, such as solvation[3,4] and proton transfer dynamics.[5,6] Over the past two decades, the development of machine-learned interatomic potentials (MLIPs) has transformed molecular dynamics (MD) simulation further by extending near-first-principles accuracy to larger systems and longer timescales.[7–10] The predictive reliability of both AIMD and MLIP-MD is determined by their underlying electronic structure description, typically provided by density functional theory (DFT).[11,12] However, DFT calculations rely on approximate exchange-correlation (XC) functionals that suffer from self-interaction and delocalization errors,[13] which can lead to quantitative inaccuracy or even complete failure qualitatively.

Correlated wavefunction (CW) theory, in contrast, provides a systematic, physically rigorous route to go beyond DFT and can provide quantitative accuracy.[14] CW methods such as second-order Møller-Plesset perturbation theory (MP2)[15] and coupled-cluster theory with single, double and perturbative triple excitations (CCSD(T))[16] reliably account for electron correlation and exact change, enabling a superior description of chemical reaction thermodynamics, especially for main group, closed-shell species.[17,18] That being said, extending CW methods to AIMD or MLIP-MD faces significant challenges.[19–21] First, the computational cost of CW methods increases steeply with system size, rendering their direct application to MD of extended systems intractable. Even with recent advances in local correlation[22,23] and periodic CW algorithms,[24] even static simulations remain limited to systems containing only a few dozen atoms -- too small to capture the structural complexity of condensed-phase chemical processes. Second, analytical energy gradients,[25,26] which are essential for MD, are often unavailable or unaffordable for CW methods. The lack of force information further hinders the training of MLIPs, which depend critically on accurate force data to describe high-dimensional energy landscapes.[27]

Many efforts have been made to overcome these challenges. From the electronic structure perspective, a variety of embedding approaches have been developed to enable application of CW theory to extended systems.[28–31] These methods assume that the chemically active region is spatially localized. The local region of interest is then treated with a high-level CW theory, while the remaining environment is described using a lower-level DFT method. Among these methods, density functional embedding theory (DFET)[32,33] and the subsequent embedded correlated wavefunction (ECW) theory,[34] which employs the electron density as the embedding variable and a uniquely defined, exact (within DFT) embedding potential to capture reactive-system-environment interactions, offer a balance between computational cost and accuracy. ECW theory has been successfully applied to many systems, including electrocatalysis,[35–40] plasmonic photocatalysis,[41,42] and aqueous-phase reactions.[43–47]

From the machine-learning perspective, transfer learning has emerged as an effective strategy for bridging different levels of electronic-structure theory while greatly reducing the amount of downstream data required.[48–50] Recently, transfer learning has been successfully applied to finetune MLIPs for small-molecule datasets[51,52] and liquid water.[53] For example, Chen *et al.* showed that using as few as 200 data points is sufficient to correct the structural properties of water, demonstrating its capability to systematically refine DFT-based MLIP models toward CW accuracy, even though their periodic CW calculations were limited to small simulation cells (16 waters).[53]

Here, we propose a framework that integrates ECW theory with transfer learning (ECW-TL), combining the efficiency of data-driven transfer-learning approaches with the quantitative accuracy of ECW methods for condensed-phase systems. We demonstrate the capability of ECW-TL by applying it to compute the free-energy profile of calcium-carbonate ion pairing in aqueous solution, a fundamental step in $CO_2$ mineralization in seawater. Using ECW-TL, we first confirm that transfer learning between two DFT approximations reproduces the target free-energy surface (FES) within chemical accuracy (~1 kcal/mol) for all critical solvation states and transition states. Then, incorporating embedded periodic MP2 and localized natural-orbital CCSD(T) data further refines the original DFT-revPBE-D3(BJ) FES, yielding a state-of-the-art MLIP that retains CW-level fidelity. Beyond ion pairing, the ECW-TL framework provides a general, efficient route toward chemically accurate simulations of reactions in complex aqueous and interfacial processes.

**ECW-TL Framework**

As summarized in Fig. 1, our ECW-TL framework consists of five stages:

(1) **Baseline model training:** Train a baseline DFT-MLIP model to accurately reproduce the DFT potential-energy surface and target properties of the system of interest, providing sufficient sampling across relevant configurations. In this work, we use the Deep Potential (DP)[54,55] framework for the MLIP model and an iterative "training - exploration - labeling" active learning procedure to explore and converge the configuration space.[56] During exploration, MLIP-MD samples new configurations using the MLIP from the previous iteration. Each configuration is evaluated by a committee of independently trained MLIPs; those with large force deviations among committee members are flagged as uncertain, labeled to perform reference DFT calculations, and added to the training set for the next iteration.

(2) **Representative subset selection:** Once the baseline DFT-MLIP model converges, select a representative subset of configurations from the training dataset. The selection can be guided by both chemical intuition and structural descriptors that characterize the atomic environments. In this work, we employ a farthest point sampling (FPS) algorithm[57] based on DP local descriptors (of the baseline model) to obtain a diverse subset. The DP local descriptor is defined as the output vector of the embedding network in a DP-MLIP model. For each configuration, we average the local descriptors over all atoms with the same element type to obtain a single configuration-level descriptor used for FPS.

(3) **ECW data generation**: For each configuration in the selected subset, consistently partition the system into a local region of interest (a "cluster") and its environment. Perform DFET/ECW calculations and collect the ECW-corrected total energies:

$$E_{\text{tot}}^{\text{ECW}} = E_{\text{tot}}^{\text{DFT}} + \left(E_{\text{emb,cluster}}^{\text{CW}} - E_{\text{emb,cluster}}^{\text{DFT}}\right), \tag{1}$$

where $E_{\text{emb,cluster}}^{\text{CW}}$ and $E_{\text{emb,cluster}}^{\text{DFT}}$ denote the energies of the embedded cluster computed at the CW and DFT levels of theory in the presence of the embedding potential derived from DFET.

(4) **Transfer learning (finetuning)**: Finetune the baseline DFT-MLIP model on the ECW corrected dataset. During transfer learning, we empirically freeze early neural network layers (i.e., the embedding network in the DP framework) and use a smaller learning rate to avoid forgetting knowledge of the pretrained DFT-based MLIP model and overfitting to the limited ECW dataset.

(5) **Validation and iterative refinement:** Run MD with the ECW-TL model to evaluate target properties. If the target property does not meet convergence or accuracy criteria, return to stage (2) and select additional configurations for DFET/ECW calculation and continue the iterative cycle until desired accuracy is achieved.

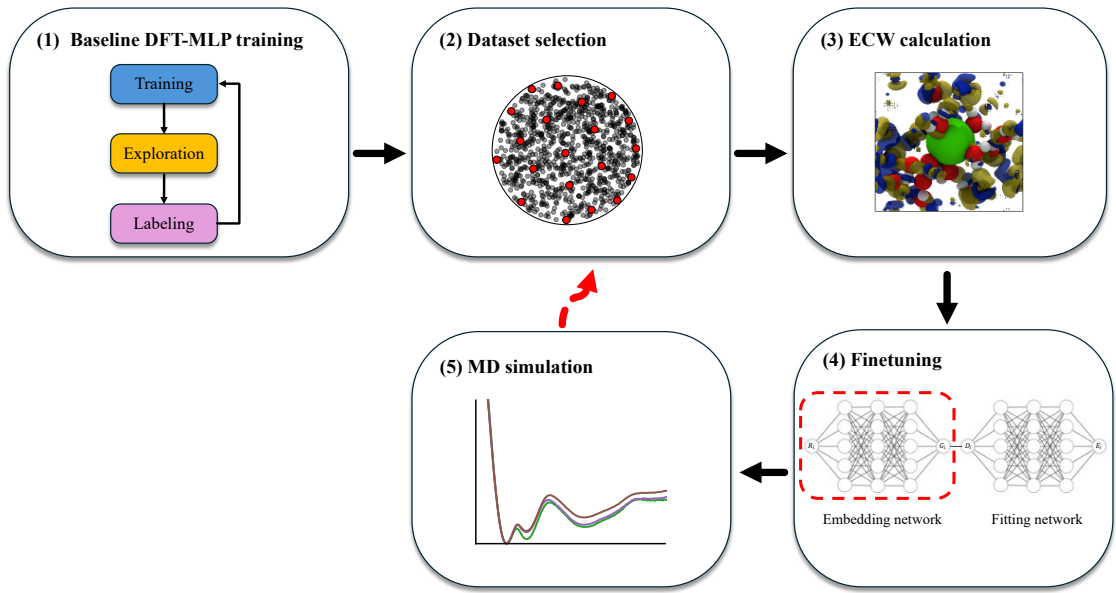

**Figure 1** Schematic workflow of the ECW-TL framework. (1) A baseline model is trained using an active-learning workflow. (2) A diverse subset of configurations is selected from the baseline dataset. (3) DFET/ECW calculations are performed on the selected configurations. (4) The model is finetuned using ECW data, during which part of the neural network (e.g., the embedding network, highlighted by the red dashed circle) is frozen to avoid overfitting. (5) MD simulations are then carried out to assess convergence. If convergence is not achieved, additional data is selected and the finetuning cycle is repeated.

Before moving to our proof-of-principle application, we highlight several key features of the ECW-TL framework. First, no force data is used during finetuning, as nuclear gradients remain difficult to compute within ECW theory. The model derives ECW-level forces from energy corrections, effectively reusing and refining the force field learned from the pretrained DFT MLIP model. Second, ECW-TL is trained entirely on condensed-phase configurations, distinguishing it from existing transfer learning or Δ-learning approaches that rely on gas-phase cluster data to extrapolate to bulk systems.[58–63] It is known that such "cluster-to-bulk" training can be problematic due to the distinct structural and electronic environments of isolated clusters versus the condensed phase.[64,65] Third, compared with directly using high-level energies for the training, the ECW-TL framework reduces size and methodological inconsistencies through the ECW energy expression in Eq. (1). The term in parentheses represents the relative energy difference between two levels of theory, effectively capturing the spirit of Δ-learning within a physically consistent embedding formulation. Together, these features make ECW-TL a practical and general framework for transferring CW accuracy into large-scale molecular simulations.

## Results

$Ca^{2+}$-$CO_3^{2-}$ ion pairing is a fundamental first step in metal-carbonate nucleation toward $CO_2$ mineralization.[66–68] Long-standing debates exist regarding whether $CaCO_3$ crystallization follows a classical nucleation pathway or proceeds through prenucleation clusters,[69–71] and obtaining a quantitatively accurate ion-pairing FES has been shown to be crucial[71] for large-scale coarse-grained simulations and understanding the microscopic mechanisms of mineral formation.

Accurately simulating $Ca^{2+}$-$CO_3^{2-}$ ion pairing is a nontrivial task. This process is governed by a subtle balance of electrostatics, polarization, and solvent rearrangement which requires a high-level electronic structure theory treatment to describe accurately. Moreover, ion association and dissociation are rare events controlled by slow solvent reorganization and collective fluctuations, which make the convergence of the FES particularly challenging. As shown in Refs.45,72–74, AIMD and MLIP-MD simulations using different electronic structure methods and enhanced-sampling schemes yield qualitatively different energetic orderings of the various solvation states. To this end, the $Ca^{2+}$-$CO_3^{2-}$ ion-pairing system provides an ideal benchmark for evaluating the accuracy, transferability, and efficiency of our ECW-TL framework.

Following our group's previous work,[45] all electronic structure calculations and MD simulations used for model training were performed using a periodic supercell containing one $Ca^{2+}$ cation, one $CO_3^{2-}$ anion, and 53 water molecules, corresponding to a ~1 M ionic concentration. This supercell is sufficiently large to capture the three key ion-pair states: bidentate and monodentate contact ion pairs (CIPs), and solvent-shared ion pairs (SSIPs). It is not large enough to study solvent-separated ion pairs.

**Validation**

Following the ECW-TL workflow, we start by training a DFT-MLIP at the revised Perdew-Burke-Ernzerhof[75,76] level of theory combined with Grimme's D3 dispersion correction with Becke-Johnson damping[77–79] (revPBE-D3(BJ)) as our baseline model. After 15 active-learning iterations, approximately 7,000 structures were collected to achieve convergence of the baseline DFT-revPBE-D3(BJ) potential. In parallel, we trained another DFT-MLIP using the strongly constrained and appropriately normed (SCAN)[80] functional to serve as our first "high-level" reference.

Enhanced-sampling MLIP-MD simulations were then performed for both models to compute the ion-pairing FES. All simulations were carried out in the NVT ensemble at 330 K, using the Ca-C distance as the only collective variable (CV). The choice of an elevated temperature reflects the upward shift in the melting temperature of water predicted by these DFT functionals.[81,82] As it happens, as shown in Fig. S4 in the Supporting Information (SI), the FES is insensitive to temperature in the range of 300-330K anyway. Previous work[74] established that a single CV is sufficient to capture the ion-pairing mechanism for this system. Further details on the DFT calculations, MLIP training, and enhanced-sampling MD simulations are provided in the Methods section below.

Next, at each ECW/embedded-DFT-SCAN-TL iteration, we selected a subset of configurations from the DFT-revPBE-D3(BJ) MLIP model training set. For each selected configuration, the cluster of interest was consistently defined as the $Ca^{2+}$ and $CO_3^{2-}$ ions together with their first solvation shell, comprising 14 water molecules as illustrated in Fig.2(b,c). This cluster size was validated in our group's earlier work.[45] Then, embedding potentials were generated using DFET (Fig.2(d)) and embedded-DFT-SCAN calculations were performed to obtain the embedded energy corrections for the cluster in the presence of its environment for these configurations. The resulting embedded-DFT-SCAN energies were used to finetune the baseline DFT-revPBE-D3(BJ) MLIP model, after which enhanced-sampling MD simulations were carried out to compute the FES with the finetuned embedded-DFT-SCAN MLIP model.

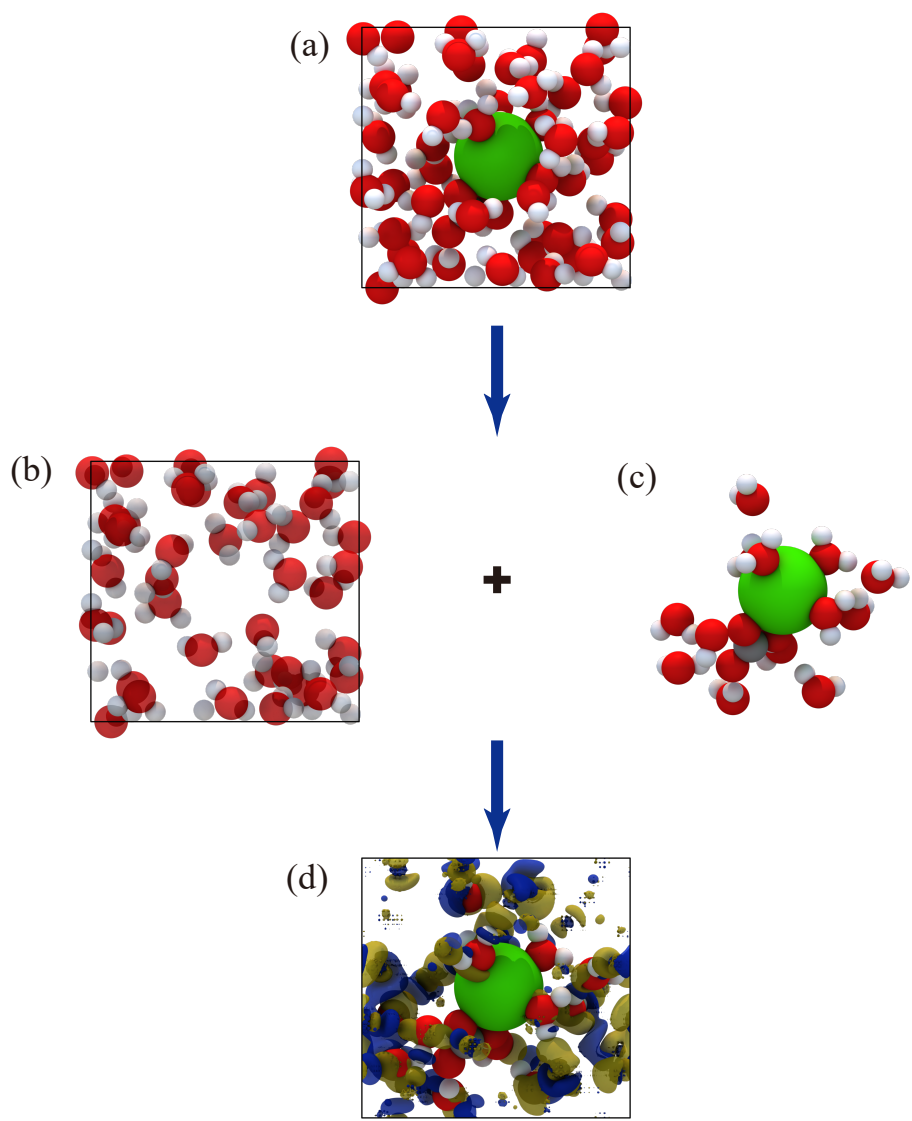

**Figure 2** Schematic illustration of the DFET/ECW calculation setup. The total simulation cell (a) is partitioned into a cluster of interest (c) and its surrounding environment (b). The CW calculation is performed on the cluster in the presence of the embedding potential (d) derived from DFET. The blue and yellow isosurfaces represent the attractive and repulsive components of the embedding potential, respectively, plotted at $V_{emb} = \pm 0.3$ V, depicting the interaction between the environment (b) and the cluster (c).

In Fig. 3a, we show the distribution of finetuning data used in each iteration, and in Fig. 3b we compare the resulting FESs obtained from the two DFT-based models (revPBE-D3(BJ) and SCAN) and the embedded-DFT-SCAN finetuned models at successive ECW-TL iterations. Comparing the two DFT models, we find the DFT-revPBE-D3(BJ) baseline model qualitatively disagrees with the DFT-SCAN model; in particular, DFT-revPBE-D3(BJ) model predicts the monodentate state to be more stable than the bidentate state. This discrepancy again highlights the need to go beyond functional-dependent DFT models to obtain a quantitatively reliable description.

We next turn to the transfer learning results. In the first iteration, approximately 700 configurations were uniformly selected along the Ca-C distance using the FPS algorithm. After the first finetuning iteration, the FES predicted by the embedded-DFT-SCAN model clearly shifts from the baseline DFT-revPBE-D3(BJ) result toward the reference DFT-SCAN result and correctly recovers the free energy ordering. However, deviations remain in the transition regions between the bidentate ($R_{Ca-C} \approx 2.8$ Å) and monodentate ($R_{Ca-C} \approx 3.3$ Å) CIP states, as well as between the monodentate CIP and the SSIP ($R_{Ca-C} \approx 4.8$ Å) states. To further improve accuracy, an additional 400 configurations from the bidentate–monodentate region were incorporated in iteration 2, which significantly refined the FES in that region. In iteration 3, another 400 configurations were added from the monodentate-SSIP region, further improving agreement with the DFT-SCAN reference.

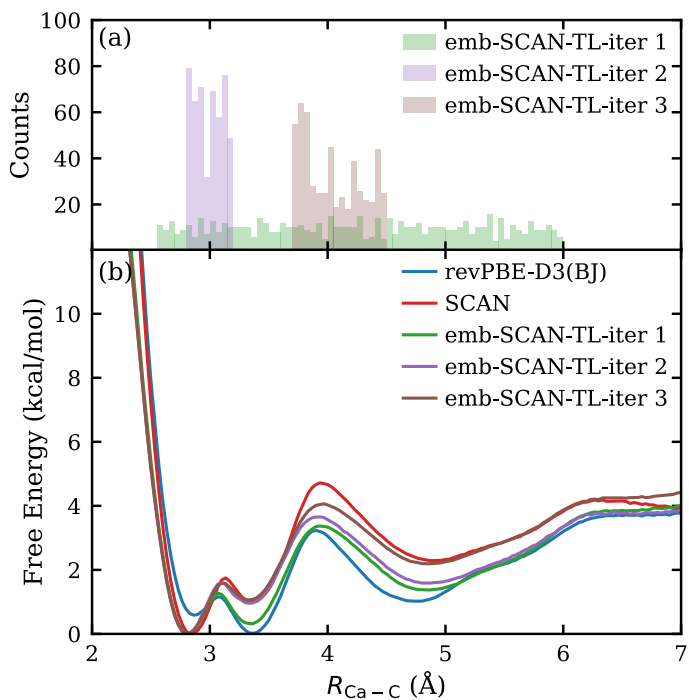

**Figure 3** (a) Distribution of the finetuning dataset along the Ca-C distance used in each ECW-TL iteration. (b) Free-energy surfaces computed from the two DFT models and the embedded-DFT-SCAN finetuned models at successive iterations. The embedded-DFT-SCAN model systematically improves throughout the ECW-TL workflow and achieves near-reference accuracy by iteration 3 across all critical solvation states.

With this finetuning dataset of ~1,500 configurations, we find the embedded-SCAN model successfully reproduces the true SCAN FES across all critical solvation states and transition states with an accuracy better than 1 kcal/mol. These results demonstrate that the selected dataset effectively samples the relevant configuration space and that the ECW-TL framework therefore should reliably transfer from DFT to high-level ECW theory with a reasonable amount of data. Moreover, for comparison we performed transfer learning using the vacuum-cluster-corrected energies (see Sec. S5 in the SI for more details) while keeping the training configurations and procedure fixed. Our ECW-TL approach is significantly more accurate than the vacuum-cluster transfer learning, highlighting the importance of using the embedding formalism. The results are shown in Fig. S8 of the SI.

**Application**

Building on the success of the DFT-level finetuning, we applied the ECW-TL framework to incorporate higher-level ECW methods. To avoid spurious cluster-image interactions arising from the finite supercell, we employed periodic Gaussian-type orbital (GTO) embedded DFT and ECW calculations, ensuring a consistent treatment of interactions between the cluster, the periodic environment, and all periodic replicas. Note that the emb-DFT-revPBE-D3(BJ) and emb-DFT-SCAN energies used for finetuning in the Validation section above also were computed within the periodic GTO formalism. This periodic approach is somewhat different from previous DFET/ECW calculations, as detailed further in the Methods section below and the SI.

Using the same structural dataset combined from the three finetuning iterations above (containing all ~ 1500 configurations), we performed periodic embedded MP2 and periodic embedded localized natural orbital CCSD(T) (LNOCCSD(T))[83] calculations and finetuned the DFT-revPBE-D3(BJ) baseline model accordingly. The resulting ECW-TL-MP2 and ECW-TL-LNOCCSD(T) free-energy surfaces are shown in Fig. 4. Remarkably, the ECW-TL-MP2 and ECW-TL-LNOCCSD(T) results are in excellent agreement with each other, consistent with previous findings[45] that MP2 provides near-quantitative accuracy for this system. In contrast, although the overall energetic ordering agrees with DFT-SCAN model, both ECW-TL-MP2 and ECW-TL-LNOCCSD(T) predict a significantly larger free-energy difference (~5 kcal/mol) between the SSIP and bidentate states compared to DFT models (~1-2 kcal/mol), showing how CW methods substantially modify the relative stability of the ion-pairing states.

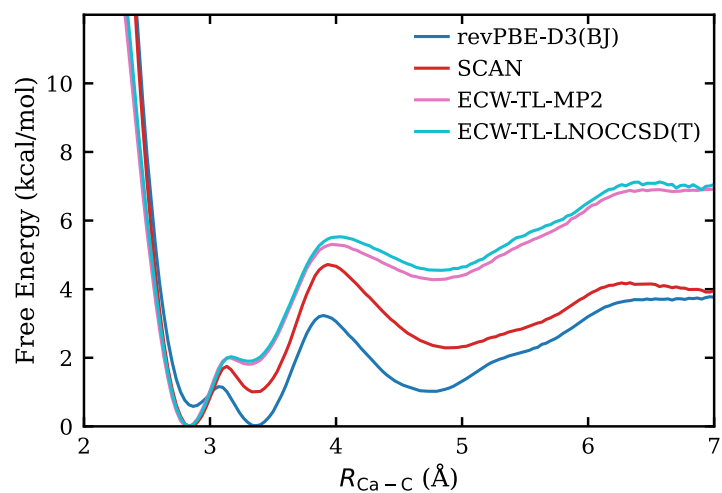

**Figure 4** Free-energy surfaces computed from the two DFT-level models and the two ECW-TL finetuned models. The two ECW-TL models agree with each other but differ qualitatively from both DFT models.

The reduction of the SSIP to CIP barrier from ~2 kcal/mol predicted by the semilocal DFT-models to ~1 kcal/mol obtained from CW-models highlights the impact of DFT delocalization error, which spuriously stabilizes charge-separated states like the SSHIP and hence overestimates the free-energy barrier for CIP formation.[73] In contrast, models based on ECW theory eliminate this artifact[45,46] and thus provide a more quantitatively accurate description of the ion-pairing energetics.

Next, we turn to the structural properties obtained from the four models discussed above. We performed constrained MD simulations at the bidentate CIP, monodentate CIP, and SSIP minima, holding the Ca-C distance fixed at each minimum while allowing all other degrees of freedom to be dynamically sampled. As shown in Fig. 5, in all three cases, the embedded-DFT-SCAN model accurately reproduces the Ca-$O_w$ (calcium - water oxygen) radial distribution function (RDF) compared to the DFT-SCAN reference and exhibits noticeable deviations from the baseline DFT-revPBE-D3(BJ) model. This demonstrates that our ECW-TL framework can capture not only the relative energetics but also the potential-energy landscape and the equilibrium structural distributions of the target high-level theory. Furthermore, DFT-SCAN, embedded DFT-SCAN finetuned and ECW-TL-LNOCCSD(T) models have a larger peak at the first solvation shell of $Ca^{2+}$ ($R_{Ca-O_w} \approx 2.4$ Å) compared to the DFT-revPBE-D3(BJ) model (insets of Fig.5), indicating a more tightly coordinated first hydration shell. This difference reflects the improved treatment of exchange and correlation in the SCAN and CW-based descriptions (more localized cation charge due to less delocalization error), which modifies the short-range, ion-water interaction potential.

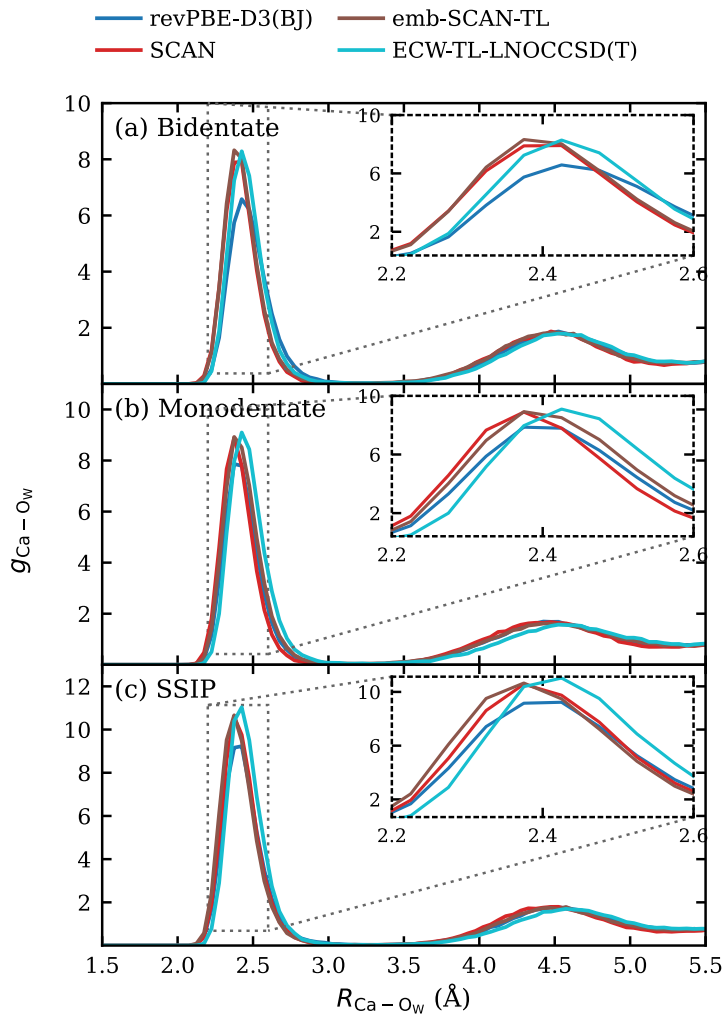

**Figure 5** Ca-O$_w$ RDFs computed from the four models at three representative Ca-C distances (a) bidentate CIP minimum ($R_{Ca-C} = 2.8$ Å), (b) monodentate CIP minimum ($R_{Ca-C} = 3.3$ Å) and (c) SSIP minimum ($R_{Ca-C} = 4.8$ Å). The finetuned emb-DFT-SCAN model agrees nearly perfectly with the full DFT-SCAN model, which is remarkable, given that no forces (only energies) from emb-DFT-SCAN were used in the finetuning. The first solvation shell structure ($R_{Ca-O_W} \approx 2.4$ Å) predicted by both finetuned models differs qualitatively from the baseline DFT-revPBE-D3(BJ) model.

In Fig. 6, we further analyze the global structural properties through the oxygen-oxygen RDF of the water molecules. Notably, all finetuned models reproduce the water structure of the baseline model rather than that of the higher-level references. This behavior is expected, as ECW corrections are applied only within the localized embedded cluster and do not affect the environmental water dynamics. Nevertheless, this level of accuracy is sufficient for our purposes, as most chemical processes involve local chemical bond changes, which the ECW-TL framework is designed to capture effectively, as shown by the ion-pairing FES.

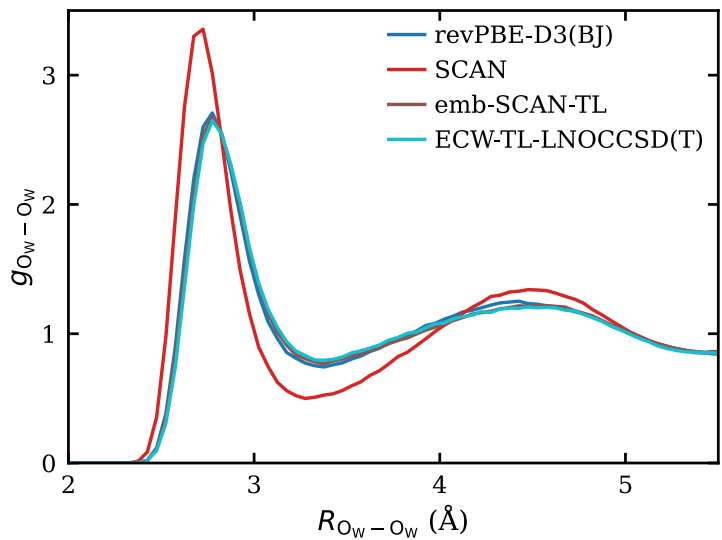

**Figure 6** $O_w$-$O_w$ RDFs computed from the four models. Neither the embedded SCAN finetuned model nor the ECW-TL-LNOCCSD(T) model alters the bulk water structure relative to the baseline revPBE-D3(BJ) models, as expected.

## Discussion and Conclusions

While these results validate the overall reliability of our ECW-TL framework, certain limitations remain. In Fig. 3(b), at the transition state between the monodentate CIP and SSIP regions ($R_{Ca-C}$ = 4.0 Å), a ~0.6 kcal/mol discrepancy exists between the third iteration embedded-DFT-SCAN and the reference DFT-SCAN results. We attempted to reduce this error by adding more configurations to the finetuning dataset but the results exhibited convergence behavior close to iteration 3, indicating that the remaining difference is likely not due to insufficient training data. This small residual error may arise from multiple sources. One possible interpretation is that, along the pathway from the SSIP to the monodentate CIP structure, a water molecule may leave the first solvation shell and diffuse into the bulk solvent, which is not accounted for within the DFET/ECW partitioning definition. Fig. 7 displays the 2D FES from the underlying DFT-revPBE-D3(BJ) theory (see Methods for details), in which both five- and six-fold coordination states are seen to be equally accessible and readily interchangeable in the CIP basin, whereas the SSIP basin is predominantly six-fold coordinated (based on the deeper blue color of the well). This suggests that CIP formation can be coupled to the release of a water molecule from the first solvation shell, among various paths for CIP formation. Accurately learning this process requires high-level information about the extended water interactions, which is difficult to include with the current embedded cluster size. Another possibility is the limited representation ability of the current MLIP model architecture. In the future, it will be valuable to explore the influence of the embedded cluster size and to test more advanced MLIP architectures, for example, multi-head architectures trained simultaneously on multiple levels of theory, or message-passing models that better capture many-body correlations.

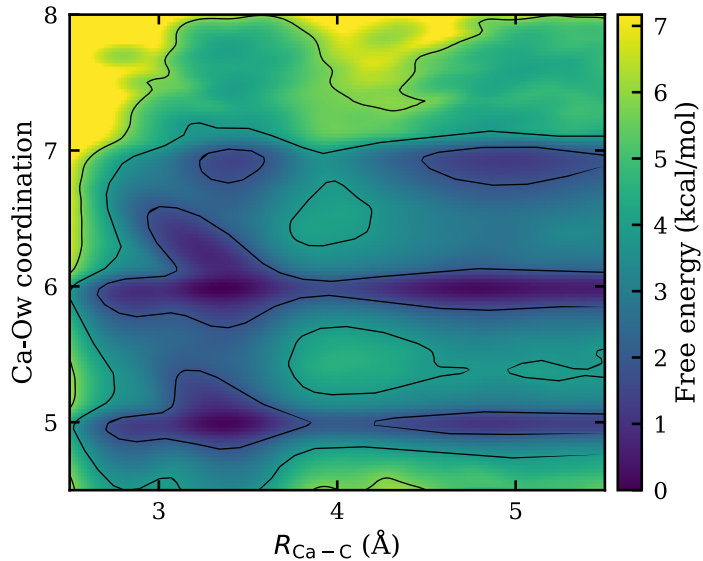

**Figure 7** Two-dimensional free-energy surface as a function of the Ca-C distance and Ca-Ow coordination number, computed from the DFT-revPBE-D3(BJ) MLIP model with the Ca-C distance as the only enhanced sampling CV. Along the pathway from the SSIP minimum to the monodentate CIP minimum, the $Ca^{2+}$ ion can adopt either five- or six-fold coordination with water molecules.

Moreover, as discussed above, the present ECW-TL framework focuses primarily on local dynamics and is thus less suited for capturing statistical observables in homogeneous systems. A potential path forward we are currently exploring is a divide-and-conquer strategy, in which a large simulation cell is partitioned into multiple clusters and ECW calculations can be performed on each cluster.

Finally, note that the ECW methods used here -- embedded MP2 and CCSD(T) -- are not well suited for electrochemical systems involving metals, and multireference approaches such as complete active space self-consistent field (CASSCF) and beyond are required. In such cases, one must ensure not only a consistent choice of embedded cluster size but also a consistent active space across different configurations,[84] aspects that we plan to address in future ECW-TL work.

In summary, the ECW-TL framework provides a practical route for transferring CW accuracy to MLIPs in the condensed phase. By combining high-level ECW corrections with systematic finetuning of DFT-MLIPs, we achieve a quantitative description of ion-pairing thermodynamics that is inaccessible to standard DFT models. The FES obtained from the
14

"gold-standard" embedded LNO-CCSD(T) calculations should approach nearly chemical accuracy for condensed-phase systems containing only closed-shell main group molecules. Future extensions incorporating multiple ion pairs and larger ionic clusters into the training set, combined with coarse-grained modeling strategies, will enable access to larger length and time scales more relevant to nucleation phenomena.[71] Given its generality, the ECW-TL approach holds great promise for direct application to a broad range of condensed-phase reactions beyond ion pairing, combining accuracy of ECW with long-time, large-scale sampling of MLIPs.

## Methods

### DFT

Periodic plane-wave DFT calculations were performed using the all-electron, frozen-core, projector augmented-wave (PAW) method within the Vienna *Ab initio* Simulation Package (VASP).[85,86] The standard PBE PAW potentials were employed for O, C, and H, and the PBE PAW potential with 3s and 3p semicore states treated as valence states was used for Ca. A kinetic-energy cutoff of 720 eV was applied, and all calculations were carried out using Γ-point k-point sampling. Gaussian smearing with a width of 0.1 eV was used to accelerate convergence, with the DFT convergence threshold set to $10^{-6}$ eV. All configurations consisted of 53 water molecules and one Ca-$CO_3$ ion pair within an orthorhombic simulation cell of dimensions 12.28 Å × 11.82 Å × 11.55 Å.

### DFET/ECW

Embedding potentials $V_{\text{emb}}(\boldsymbol{r})$ on real space grids within the periodic cell were generated at the DFT-revPBE-D3(BJ) level in a modified version of VASP[87] via PAW-DFET within a planewave basis by maximizing an extended Wu-Yang functional[88,89]:

$$W = E_{\text{cluster}}^{\text{DFT}}[\rho_{\text{cluster}}, V_{\text{emb}}] + E_{\text{enviro}}^{\text{DFT}}[\rho_{\text{enviro}}, V_{\text{emb}}] - \int d\boldsymbol{r}\, V_{\text{emb}}(\boldsymbol{r})\, \rho_{\text{tot}}(\boldsymbol{r}), \qquad (2)$$

where: $E_{\text{cluster}}^{\text{DFT}}$ and $E_{\text{enviro}}^{\text{DFT}}$ are self-consistent DFT energies of the cluster and its environment subject to the added external potential, $V_{\text{emb}}$; $\rho_{\text{cluster}}$ and $\rho_{\text{enviro}}$ represent the corresponding electron densities; and convergence is reached within the DFET when the total electron density from the entire periodic cell $\rho_{\text{tot}} = \rho_{\text{cluster}} + \rho_{\text{enviro}}$. For additional details, we refer the reader to Refs.33,41 and 87.

The embedding potentials on the real-space grid then were projected onto a cc-pVTZ[90] basis set for use in our periodic embedded calculations.[91] Embedded periodic GTO DFT-revPBE-D3(BJ), DFT-SCAN and ECW calculations then were performed using the PySCF package.[92] Embedded periodic planewave DFT and embedded periodic GTO DFT total energies agree quite well with each other (Fig. S6), validating our choice to perform embedded periodic GTO calculations for both parts of the regional correction term in Eq. (1) for basis set consistency, and thereby capturing both periodic image interactions and enabling periodic CW calculations in the regional correction term. The periodic LNO-CCSD(T) calculations were performed using a separate branch of PySCF.[93] The final LNO-CCSD(T) energies were augmented with a ΔMP2 correction for improved convergence, following the procedure described in Ref. [83]. For more computational details and convergence tests (Fig. S5 and S7) on periodic GTO calculations, see the SI.

**MLIP**

MLIPs were trained using the DeePMD-kit package[94] together with an active-learning procedure implemented in the DPGEN package.[56] All models used the standard short-range, smooth two-body descriptor with a cutoff radius of $r_c = 6$ Å. We used the "finetune" function in DeePMD-kit for transfer learning and froze the embedding network during finetuning. For additional details on the training procedure, validation (Fig. S1 and S2), finetuning protocol, and assessment of the MLIP models, see the SI.

**MD**

The FESs shown in the main text were computed using on-the-fly probability-enhanced sampling (OPES)[95] implemented in the LAMMPS package[96] with the PLUMED plugin.[97] Each FES was obtained by averaging four FESs calculated from four independent 20-ns trajectories to ensure sufficient sampling. The Ca-Ow coordination number was defined according to a cubic switching function[97] with $D_{max}$= 3.75 Å and $D_0$=2.5 Å. More details are provided in the SI, where we also compare results with the constrained reaction coordinate, blue-moon ensemble (BME) MD method;[98,99] the two methods exhibit excellent agreement with each other (Fig. S3).

The training data, ML models, and all input scripts to reproduce the simulations are available at https://github.com/xzbian/ECW-TL-CaCO3.

**Note added after submission.** After submitting this manuscript, we became aware of a related study by O'Neil *et al.*,[100] which investigates $Ca^{2+}$-$CO_3^2$ ion pairing using a cluster-to-bulk Δ-learning strategy to improve machine-learning potential with correlated wavefunction information.

**Acknowledgements**

This work was supported by the Computational Chemical Science Center: Chemistry in Solution and at Interfaces, supported as part of the Computational Chemical Sciences Program funded by the U.S. Department of Energy (DOE), Office of Science, Basic Energy Sciences, under Award No. DESC0019394. X.B. was also partially supported by the BP Carbon Mitigation Initiative (CMI) and the Princeton Catalysis Initiative at Princeton University. We acknowledge the computational resources provided by the National Energy Research Scientific Computing Center (NERSC), which is supported by the DOE Office of Science under Contract No. DE-AC0205CH11231, and Princeton Research Computing at Princeton University. X.B. thanks Pablo Piaggi, Hong-zhou Ye, Lingqing Peng, Ziyang Wei and John Mark Martirez for helpful discussions.

# References

(1) Car, R.; Parrinello, M. Unified Approach for Molecular Dynamics and Density-Functional Theory. *Phys. Rev. Lett.* **1985**, *55* (22), 2471–2474. https://doi.org/10.1103/PhysRevLett.55.2471.

(2) Marx, D.; Hutter, J. *Ab Initio Molecular Dynamics: Basic Theory and Advanced Methods*; Cambridge University Press, 2009.

(3) Tuckerman, M.; Laasonen, K.; Sprik, M.; Parrinello, M. Ab Initio Molecular Dynamics Simulation of the Solvation and Transport of Hydronium and Hydroxyl Ions in Water. *J. Chem. Phys.* **1995**, *103* (1), 150–161. https://doi.org/10.1063/1.469654.

(4) Ramaniah, L. M.; Bernasconi, M.; Parrinello, M. Ab Initio Molecular-Dynamics Simulation of K+ Solvation in Water. *J. Chem. Phys.* **1999**, *111* (4), 1587–1591. https://doi.org/10.1063/1.479418.

(5) Geissler, P. L.; Dellago, C.; Chandler, D.; Hutter, J.; Parrinello, M. Ab Initio Analysis of Proton Transfer Dynamics in (H2O)3H+. *Chem. Phys. Lett.* **2000**, *321* (3), 225–230. https://doi.org/10.1016/S0009-2614(00)00381-X.

(6) Marx, D. Proton Transfer 200 Years after von Grotthuss: Insights from Ab Initio Simulations. *Chemphyschem Eur. J. Chem. Phys. Phys. Chem.* **2006**, *7* (9), 1848–1870. https://doi.org/10.1002/cphc.200600128.

(7) Behler, J.; Parrinello, M. Generalized Neural-Network Representation of High-Dimensional Potential-Energy Surfaces. *Phys. Rev. Lett.* **2007**, *98* (14), 146401. https://doi.org/10.1103/PhysRevLett.98.146401.

(8) Behler, J. Perspective: Machine Learning Potentials for Atomistic Simulations. *J. Chem. Phys.* **2016**, *145* (17), 170901. https://doi.org/10.1063/1.4966192.

(9) Noé, F.; Tkatchenko, A.; Müller, K.-R.; Clementi, C. Machine Learning for Molecular Simulation. *Annu. Rev. Phys. Chem.* **2020**, *71* (Volume 71, 2020), 361–390. https://doi.org/10.1146/annurev-physchem-042018-052331.

(10) Unke, O. T.; Chmiela, S.; Sauceda, H. E.; Gastegger, M.; Poltavsky, I.; Schütt, K. T.; Tkatchenko, A.; Müller, K.-R. Machine Learning Force Fields. *Chem. Rev.* **2021**, *121* (16), 10142–10186. https://doi.org/10.1021/acs.chemrev.0c01111.

(11) Hohenberg, P.; Kohn, W. Inhomogeneous Electron Gas. *Phys. Rev.* **1964**, *136* (3B), B864–B871. https://doi.org/10.1103/PhysRev.136.B864.

(12) Kohn, W.; Sham, L. J. Self-Consistent Equations Including Exchange and Correlation Effects. *Phys. Rev.* **1965**, *140* (4A), A1133–A1138. https://doi.org/10.1103/PhysRev.140.A1133.

(13) Cohen, A. J.; Mori-Sánchez, P.; Yang, W. Insights into Current Limitations of Density Functional Theory. *Science* **2008**, *321* (5890), 792–794. https://doi.org/10.1126/science.1158722.

(14) Helgaker, T.; Jorgensen, P.; Olsen, J. *Molecular Electronic-Structure Theory*; John Wiley & Sons, 2013.

(15) Møller, Chr.; Plesset, M. S. Note on an Approximation Treatment for Many-Electron Systems. *Phys. Rev.* **1934**, *46* (7), 618–622. https://doi.org/10.1103/PhysRev.46.618.

(16) Raghavachari, K.; Trucks, G. W.; Pople, J. A.; Head-Gordon, M. A Fifth-Order Perturbation Comparison of Electron Correlation Theories. *Chem. Phys. Lett.* **1989**, *157* (6), 479–483. https://doi.org/10.1016/S0009-2614(89)87395-6.

(17) Stanton, J. F. Why CCSD(T) Works: A Different Perspective. *Chem. Phys. Lett.* **1997**, *281* (1), 130–134. https://doi.org/10.1016/S0009-2614(97)01144-5.

(18) Fink, R. F. Why Does MP2 Work? *J. Chem. Phys.* **2016**, *145* (18), 184101. https://doi.org/10.1063/1.4966689.

(19) da Silva, A. J. R.; Cheng, H.-Y.; Gibson, D. A.; Sorge, K. L.; Liu, Z.; Carter, E. A. Limitations of Ab Initio Molecular Dynamics Simulations of Simple Reactions: F + H2 as a Prototype. *Spectrochim. Acta. A. Mol. Biomol. Spectrosc.* **1997**, *53* (8), 1285–1299. https://doi.org/10.1016/S1386-1425(97)89474-7.

(20) Hayes, R. L.; Fattal, E.; Govind, N.; Carter, E. A. Long Live Vinylidene! A New View of the H2CC: → HC:CH Rearrangement from Ab Initio Molecular Dynamics. *J. Am. Chem. Soc.* **2001**, *123* (4), 641–657. https://doi.org/10.1021/ja000907x.

(21) Nandi, A.; Qu, C.; Houston, P. L.; Conte, R.; Bowman, J. M. Δ-Machine Learning for Potential Energy Surfaces: A PIP Approach to Bring a DFT-Based PES to CCSD(T) Level of Theory. *J. Chem. Phys.* **2021**, *154* (5), 051102. https://doi.org/10.1063/5.0038301.

(22) Rolik, Z.; Szegedy, L.; Ladjánszki, I.; Ladóczki, B.; Kállay, M. An Efficient Linear-Scaling CCSD(T) Method Based on Local Natural Orbitals. *J. Chem. Phys.* **2013**, *139* (9), 094105. https://doi.org/10.1063/1.4819401.

(23) Guo, Y.; Riplinger, C.; Becker, U.; Liakos, D. G.; Minenkov, Y.; Cavallo, L.; Neese, F. Communication: An Improved Linear Scaling Perturbative Triples Correction for the Domain Based Local Pair-Natural Orbital Based Singles and Doubles Coupled Cluster Method [DLPNO-CCSD(T)]. *J. Chem. Phys.* **2018**, *148* (1), 011101. https://doi.org/10.1063/1.5011798.

(24) Booth, G. H.; Tsatsoulis, T.; Chan, G. K.-L.; Grüneis, A. From Plane Waves to Local Gaussians for the Simulation of Correlated Periodic Systems. *J. Chem. Phys.* **2016**, *145* (8), 084111. https://doi.org/10.1063/1.4961301.

(25) Pople, J. A.; Krishnan, R.; Schlegel, H. B.; Binkley, J. S. Derivative Studies in Hartree-Fock and Møller-Plesset Theories. *Int. J. Quantum Chem.* **1979**, *16* (S13), 225–241. https://doi.org/10.1002/qua.560160825.

(26) Harding, M. E.; Metzroth, T.; Gauss, J.; Auer, A. A. Parallel Calculation of CCSD and CCSD(T) Analytic First and Second Derivatives. *J. Chem. Theory Comput.* **2008**, *4* (1), 64–74. https://doi.org/10.1021/ct700152c.

(27) Tokita, A. M.; Behler, J. How to Train a Neural Network Potential. *J. Chem. Phys.* **2023**, *159* (12), 121501. https://doi.org/10.1063/5.0160326.

(28) Huang, P.; Carter, E. A. Advances in Correlated Electronic Structure Methods for Solids, Surfaces, and Nanostructures. *Annu. Rev. Phys. Chem.* **2008**, *59* (Volume 59, 2008), 261–290. https://doi.org/10.1146/annurev.physchem.59.032607.093528.

(29) Chung, L. W.; Sameera, W. M. C.; Ramozzi, R.; Page, A. J.; Hatanaka, M.; Petrova, G. P.; Harris, T. V.; Li, X.; Ke, Z.; Liu, F.; Li, H.-B.; Ding, L.; Morokuma, K. The ONIOM Method and Its Applications. *Chem. Rev.* **2015**, *115* (12), 5678–5796. https://doi.org/10.1021/cr5004419.

(30) Sun, Q.; Chan, G. K.-L. Quantum Embedding Theories. *Acc. Chem. Res.* **2016**, *49* (12), 2705–2712. https://doi.org/10.1021/acs.accounts.6b00356.

(31) Yu, K.; Krauter, C. M.; Dieterich, J. M.; Carter, E. A. Density and Potential Functional Embedding: Theory and Practice. In *Fragmentation*; John Wiley & Sons, Ltd, 2017; pp 81–117. https://doi.org/10.1002/9781119129271.ch2.

(32) Govind, N.; Wang, Y. A.; da Silva, A. J. R.; Carter, E. A. Accurate Ab Initio Energetics of Extended Systems via Explicit Correlation Embedded in a Density Functional Environment. *Chem. Phys. Lett.* **1998**, *295* (1), 129–134. https://doi.org/10.1016/S0009-2614(98)00939-7.

(33) Huang, C.; Pavone, M.; Carter, E. A. Quantum Mechanical Embedding Theory Based on a Unique Embedding Potential. *J. Chem. Phys.* **2011**, *134* (15), 154110. https://doi.org/10.1063/1.3577516.

(34) Libisch, F.; Huang, C.; Carter, E. A. Embedded Correlated Wavefunction Schemes: Theory and Applications. *Acc. Chem. Res.* **2014**, *47* (9), 2768–2775. https://doi.org/10.1021/ar500086h.

(35) Zhao, Q.; Martirez, J. M. P.; Carter, E. A. Revisiting Understanding of Electrochemical $CO_2$ Reduction on Cu(111): Competing Proton-Coupled Electron Transfer Reaction Mechanisms Revealed by Embedded Correlated Wavefunction Theory. *J. Am. Chem. Soc.* **2021**, *143* (16), 6152–6164. https://doi.org/10.1021/jacs.1c00880.

(36) Zhao, Q.; Martirez, J. M. P.; Carter, E. A. Charting C–C Coupling Pathways in Electrochemical $CO_2$ Reduction on Cu(111) Using Embedded Correlated Wavefunction Theory. *Proc. Natl. Acad. Sci.* **2022**, *119* (44), e2202931119. https://doi.org/10.1073/pnas.2202931119.

(37) Zhao, Q.; Martirez, J. M. P.; Carter, E. A. Electrochemical Hydrogenation of CO on Cu(100): Insights from Accurate Multiconfigurational Wavefunction Methods. *J. Phys. Chem. Lett.* **2022**, *13* (44), 10282–10290. https://doi.org/10.1021/acs.jpclett.2c02444.

(38) Cai, J.; Zhao, Q.; Hsu, W.-Y.; Choi, C.; Liu, Y.; Martirez, J. M. P.; Chen, C.; Huang, J.; Carter, E. A.; Huang, Y. Highly Selective Electrochemical Reduction of $CO_2$ into Methane on Nanotwinned Cu. *J. Am. Chem. Soc.* **2023**, *145* (16), 9136–9143. https://doi.org/10.1021/jacs.3c00847.

(39) Martirez, J. M. P.; Carter, E. A. C–C Bond Formation during Electrochemical CO2 Reduction on Pristine Cu(100) Unlikely to Involve Adsorbed CO at Any Potential. *J. Am. Chem. Soc.* **2026**, *148* (7), 7415–7425. https://doi.org/10.1021/jacs.5c20300.

(40) Martirez, J. M. P.; Carter, E. A. Insights into Nonelectroactive C–C Bond Formation on Cu(100) during Electrochemical CO2 Reduction from Multiconfigurational Wavefunction Theory. *J. Phys. Chem. C* **2026**. https://doi.org/10.1021/acs.jpcc.5c07792.

(41) Martirez, J. M. P.; Bao, J. L.; Carter, E. A. First-Principles Insights into Plasmon-Induced Catalysis. *Annu. Rev. Phys. Chem.* **2021**, *72* (1), 99–119. https://doi.org/10.1146/annurev-physchem-061020-053501.

(42) Wen, X.; Martirez, J. M. P.; Carter, E. A. Plasmon-Driven Ammonia Decomposition on Pd(111): Hole Transfer's Role in Changing Rate-Limiting Steps. *ACS Catal.* **2024**, *14* (12), 9539–9553. https://doi.org/10.1021/acscatal.4c01869.

(43) Martirez, J. M. P.; Carter, E. A. Solvent Dynamics Are Critical to Understanding Carbon Dioxide Dissolution and Hydration in Water. *J. Am. Chem. Soc.* **2023**, *145* (23), 12561–12575. https://doi.org/10.1021/jacs.3c01283.

(44) Boyn, J.-N.; Carter, E. A. Probing pH-Dependent Dehydration Dynamics of Mg and Ca Cations in Aqueous Solutions with Multi-Level Quantum Mechanics/Molecular Dynamics Simulations. *J. Am. Chem. Soc.* **2023**, *145* (37), 20462–20472. https://doi.org/10.1021/jacs.3c06182.

(45) Boyn, J.-N.; Carter, E. A. Characterizing the Mechanisms of Ca and Mg Carbonate Ion-Pair Formation with Multi-Level Molecular Dynamics/Quantum Mechanics Simulations. *J. Phys. Chem. B* **2023**, *127* (50), 10824–10832. https://doi.org/10.1021/acs.jpcb.3c05369.

(46) Boyn, J.-N.; Carter, E. A. Elucidating and Contrasting the Mechanisms for Mg and Ca Sulfate Ion-Pair Formation with Multi-Level Embedded Quantum Mechanics/Molecular Dynamics Simulations. *J. Chem. Phys.* **2024**, *161* (22), 224501. https://doi.org/10.1063/5.0235460.

(47) Bobell, B.; Boyn, J.-N.; Martirez, J. M. P.; Carter, E. A. Modeling Bicarbonate Formation in an Alkaline Solution with Multi-Level Quantum Mechanics/Molecular Dynamics Simulations. *Mol. Phys.* **2025**, *123* (7–8), e2375370. https://doi.org/10.1080/00268976.2024.2375370.

(48) Zhang, D.; Liu, X.; Zhang, X.; Zhang, C.; Cai, C.; Bi, H.; Du, Y.; Qin, X.; Peng, A.; Huang, J.; Li, B.; Shan, Y.; Zeng, J.; Zhang, Y.; Liu, S.; Li, Y.; Chang, J.; Wang, X.; Zhou, S.; Liu, J.; Luo, X.; Wang, Z.; Jiang, W.; Wu, J.; Yang, Y.; Yang, J.; Yang, M.; Gong, F.-Q.; Zhang, L.; Shi, M.; Dai, F.-Z.; York, D. M.; Liu, S.; Zhu, T.; Zhong, Z.; Lv, J.; Cheng, J.; Jia, W.; Chen, M.; Ke, G.; E, W.; Zhang, L.; Wang, H. DPA-2: A Large Atomic Model as a Multi-Task Learner. *Npj Comput. Mater.* **2024**, *10* (1), 293. https://doi.org/10.1038/s41524-024-01493-2.

(49) Niblett, S. P.; Kourtis, P.; Magdău, I.-B.; Grey, C. P.; Csányi, G. Transferability of Datasets between Machine-Learning Interaction Potentials. arXiv September 9, 2024. https://doi.org/10.48550/arXiv.2409.05590.

(50) Röcken, S.; Zavadlav, J. Enhancing Machine Learning Potentials through Transfer Learning across Chemical Elements. *J. Chem. Inf. Model.* **2025**, *65* (14), 7406–7414. https://doi.org/10.1021/acs.jcim.5c00293.

(51) Smith, J. S.; Nebgen, B. T.; Zubatyuk, R.; Lubbers, N.; Devereux, C.; Barros, K.; Tretiak, S.; Isayev, O.; Roitberg, A. E. Approaching Coupled Cluster Accuracy with a General-Purpose Neural Network Potential through Transfer Learning. *Nat. Commun.* **2019**, *10* (1), 2903. https://doi.org/10.1038/s41467-019-10827-4.

(52) Käser, S.; Boittier, E. D.; Upadhyay, M.; Meuwly, M. Transfer Learning to CCSD(T): Accurate Anharmonic Frequencies from Machine Learning Models. *J. Chem. Theory Comput.* **2021**, *17* (6), 3687–3699. https://doi.org/10.1021/acs.jctc.1c00249.

(53) Chen, M. S.; Lee, J.; Ye, H.-Z.; Berkelbach, T. C.; Reichman, D. R.; Markland, T. E. Data-Efficient Machine Learning Potentials from Transfer Learning of Periodic Correlated Electronic Structure Methods: Liquid Water at AFQMC, CCSD, and CCSD(T) Accuracy. *J. Chem. Theory Comput.* **2023**, *19* (14), 4510–4519. https://doi.org/10.1021/acs.jctc.2c01203.

(54) Zhang, L.; Han, J.; Wang, H.; Car, R.; E, W. Deep Potential Molecular Dynamics: A Scalable Model with the Accuracy of Quantum Mechanics. *Phys. Rev. Lett.* **2018**, *120* (14), 143001. https://doi.org/10.1103/PhysRevLett.120.143001.

(55) Zhang, L.; Han, J.; Wang, H.; Saidi, W.; Car, R.; E, W. End-to-End Symmetry Preserving Inter-Atomic Potential Energy Model for Finite and Extended Systems. In *Advances in Neural Information Processing Systems*; Curran Associates, Inc., 2018; Vol. 31.

(56) Zhang, Y.; Wang, H.; Chen, W.; Zeng, J.; Zhang, L.; Wang, H.; E, W. DP-GEN: A Concurrent Learning Platform for the Generation of Reliable Deep Learning Based Potential Energy Models. *Comput. Phys. Commun.* **2020**, *253*, 107206. https://doi.org/10.1016/j.cpc.2020.107206.

(57) Ceriotti, M.; Tribello, G. A.; Parrinello, M. Demonstrating the Transferability and the Descriptive Power of Sketch-Map. *J. Chem. Theory Comput.* **2013**, *9* (3), 1521–1532. https://doi.org/10.1021/ct3010563.

(58) Daru, J.; Forbert, H.; Behler, J.; Marx, D. Coupled Cluster Molecular Dynamics of Condensed Phase Systems Enabled by Machine Learning Potentials: Liquid Water Benchmark. *Phys. Rev. Lett.* **2022**, *129* (22), 226001. https://doi.org/10.1103/PhysRevLett.129.226001.

(59) Ruth, M.; Gerbig, D.; Schreiner, P. R. Machine Learning of Coupled Cluster (T)-Energy Corrections via Delta (Δ)-Learning. *J. Chem. Theory Comput.* **2022**, *18* (8), 4846–4855. https://doi.org/10.1021/acs.jctc.2c00501.

(60) Qu, C.; Yu, Q.; Conte, R.; L. Houston, P.; Nandi, A.; M. Bomwan, J. A Δ-Machine Learning Approach for Force Fields, Illustrated by a CCSD(T) 4-Body Correction to the MB-Pol Water Potential. *Digit. Discov.* **2022**, *1* (5), 658–664. https://doi.org/10.1039/D2DD00057A.

(61) Mészáros, B. B.; Szabó, A.; Daru, J. Short-Range Δ-Machine Learning: A Cost-Efficient Strategy to Transfer Chemical Accuracy to Condensed Phase Systems. *J. Chem. Theory Comput.* **2025**, *21* (11), 5372–5381. https://doi.org/10.1021/acs.jctc.5c00367.

(62) O'Neill, N.; Shi, B. X.; Baldwin, W. J.; Witt, W. C.; Csányi, G.; Gale, J. D.; Michaelides, A.; Schran, C. Towards Routine Condensed Phase Simulations with Delta-Learned Coupled Cluster Accuracy: Application to Liquid Water. *J. Chem. Theory Comput.* **2025**, *21* (22), 11710–11720. https://doi.org/10.1021/acs.jctc.5c01377.

(63) Boittier, E. D.; Käser, S.; Meuwly, M. Roadmap to CCSD(T)-Quality Machine-Learned Potentials for Condensed Phase Simulations. *J. Chem. Theory Comput.* **2025**, *21* (18), 8683–8698. https://doi.org/10.1021/acs.jctc.5c01085.

(64) Zaverkin, V.; Holzmüller, D.; Schuldt, R.; Kästner, J. Predicting Properties of Periodic Systems from Cluster Data: A Case Study of Liquid Water. *J. Chem. Phys.* **2022**, *156* (11), 114103. https://doi.org/10.1063/5.0078983.

(65) Gawkowski, M. J.; Li, M.; Shi, B. X.; Kapil, V. The Good, the Bad, and the Ugly of Atomistic Learning for "Clusters-to-Bulk" Generalization. *Mach. Learn. Sci. Technol.* **2026**. https://doi.org/10.1088/2632-2153/ae3c57.

(66) Morse, J. W.; Arvidson, R. S. The Dissolution Kinetics of Major Sedimentary Carbonate Minerals. *Earth-Sci. Rev.* **2002**, *58* (1), 51–84. https://doi.org/10.1016/S0012-8252(01)00083-6.

(67) Kellermeier, M.; Raiteri, P.; Berg, J. K.; Kempter, A.; Gale, J. D.; Gebauer, D. Entropy Drives Calcium Carbonate Ion Association. *ChemPhysChem* **2016**, *17* (21), 3535–3541. https://doi.org/10.1002/cphc.201600653.

(68) Chang, R.; Kim, S.; Lee, S.; Choi, S.; Kim, M.; Park, Y. Calcium Carbonate Precipitation for CO2 Storage and Utilization: A Review of the Carbonate Crystallization and Polymorphism. *Front. Energy Res.* **2017**, *5*. https://doi.org/10.3389/fenrg.2017.00017.

(69) Gebauer, D.; Völkel, A.; Cölfen, H. Stable Prenucleation Calcium Carbonate Clusters. *Science* **2008**, *322* (5909), 1819–1822. https://doi.org/10.1126/science.1164271.

(70) Raiteri, P.; Gale, J. D. Water Is the Key to Nonclassical Nucleation of Amorphous Calcium Carbonate. *J. Am. Chem. Soc.* **2010**, *132* (49), 17623–17634. https://doi.org/10.1021/ja108508k.

(71) Henzler, K.; Fetisov, E. O.; Galib, M.; Baer, M. D.; Legg, B. A.; Borca, C.; Xto, J. M.; Pin, S.; Fulton, J. L.; Schenter, G. K.; Govind, N.; Siepmann, J. I.; Mundy, C. J.; Huthwelker, T.; De Yoreo, J. J. Supersaturated Calcium Carbonate Solutions Are Classical. *Sci. Adv.* **2018**, *4* (1), eaao6283. https://doi.org/10.1126/sciadv.aao6283.

(72) Raiteri, P.; Demichelis, R.; Gale, J. D. Thermodynamically Consistent Force Field for Molecular Dynamics Simulations of Alkaline-Earth Carbonates and Their Aqueous Speciation. *J. Phys. Chem. C* **2015**, *119* (43), 24447–24458. https://doi.org/10.1021/acs.jpcc.5b07532.

(73) Raiteri, P.; Schuitemaker, A.; Gale, J. D. Ion Pairing and Multiple Ion Binding in Calcium Carbonate Solutions Based on a Polarizable AMOEBA Force Field and Ab Initio Molecular Dynamics. *J. Phys. Chem. B* **2020**, *124* (17), 3568–3582. https://doi.org/10.1021/acs.jpcb.0c01582.

(74) Piaggi, P. M.; Gale, J. D.; Raiteri, P. Ab Initio Machine-Learning Simulation of Calcium Carbonate from Aqueous Solutions to the Solid State. *Proc. Natl. Acad. Sci.* **2025**, *122* (41), e2415663122. https://doi.org/10.1073/pnas.2415663122.

(75) Perdew, J. P.; Burke, K.; Ernzerhof, M. Generalized Gradient Approximation Made Simple. *Phys. Rev. Lett.* **1996**, *77* (18), 3865–3868. https://doi.org/10.1103/PhysRevLett.77.3865.

(76) Zhang, Y.; Yang, W. Comment on ``Generalized Gradient Approximation Made Simple''. *Phys. Rev. Lett.* **1998**, *80* (4), 890–890. https://doi.org/10.1103/PhysRevLett.80.890.

(77) Becke, A. D.; Johnson, E. R. A Density-Functional Model of the Dispersion Interaction. *J. Chem. Phys.* **2005**, *123* (15), 154101. https://doi.org/10.1063/1.2065267.

(78) Grimme, S.; Antony, J.; Ehrlich, S.; Krieg, H. A Consistent and Accurate Ab Initio Parametrization of Density Functional Dispersion Correction (DFT-D) for the 94 Elements H-Pu. *J. Chem. Phys.* **2010**, *132* (15), 154104. https://doi.org/10.1063/1.3382344.

(79) Grimme, S.; Ehrlich, S.; Goerigk, L. Effect of the Damping Function in Dispersion Corrected Density Functional Theory. *J. Comput. Chem.* **2011**, *32* (7), 1456–1465. https://doi.org/10.1002/jcc.21759.

(80) Sun, J.; Ruzsinszky, A.; Perdew, J. P. Strongly Constrained and Appropriately Normed Semilocal Density Functional. *Phys. Rev. Lett.* **2015**, *115* (3), 036402. https://doi.org/10.1103/PhysRevLett.115.036402.

(81) Montero de Hijes, P.; Dellago, C.; Jinnouchi, R.; Kresse, G. Density Isobar of Water and Melting Temperature of Ice: Assessing Common Density Functionals. *J. Chem. Phys.* **2024**, *161* (13), 131102. https://doi.org/10.1063/5.0227514.

(82) Li, Y.; Yang, B.; Zhang, C.; Gomez, A.; Xie, P.; Chen, Y.; Piaggi, P. M.; Car, R. Assessment of First-Principles Methods in Modeling the Melting Properties of Water. arXiv December 30, 2025. https://doi.org/10.48550/arXiv.2512.23940.

(83) Ye, H.-Z.; Berkelbach, T. C. Periodic Local Coupled-Cluster Theory for Insulators and Metals. *J. Chem. Theory Comput.* **2024**, *20* (20), 8948–8959. https://doi.org/10.1021/acs.jctc.4c00936.

(84) Wen, X.; Boyn, J.-N.; Martirez, J. M. P.; Zhao, Q.; Carter, E. A. Strategies to Obtain Reliable Energy Landscapes from Embedded Multireference Correlated Wavefunction Methods for Surface Reactions. *J. Chem. Theory Comput.* **2024**, *20* (14), 6037–6048. https://doi.org/10.1021/acs.jctc.4c00558.

(85) Kresse, G.; Joubert, D. From Ultrasoft Pseudopotentials to the Projector Augmented-Wave Method. *Phys. Rev. B* **1999**, *59* (3), 1758–1775. https://doi.org/10.1103/PhysRevB.59.1758.

(86) Kresse, G.; Hafner, J. Ab Initio Molecular Dynamics for Liquid Metals. *Phys. Rev. B* **1993**, *47* (1), 558–561. https://doi.org/10.1103/PhysRevB.47.558.

(87) Kuang, Y.; Carter, E. A. VASPEmbedding. https://github.com/EACcodes/VASPEmbedding.

(88) Wu, Q.; Yang, W. A Direct Optimization Method for Calculating Density Functionals and Exchange–Correlation Potentials from Electron Densities. *J. Chem. Phys.* **2003**, *118* (6), 2498–2509. https://doi.org/10.1063/1.1535422.

(89) Yu, K.; Libisch, F.; Carter, E. A. Implementation of Density Functional Embedding Theory within the Projector-Augmented-Wave Method and Applications to Semiconductor Defect States. *J. Chem. Phys.* **2015**, *143* (10), 102806. https://doi.org/10.1063/1.4922260.

(90) Dunning, T. H. Gaussian Basis Sets for Use in Correlated Molecular Calculations. I. The Atoms Boron through Neon and Hydrogen. *J. Chem. Phys.* **1989**, *90* (2), 1007–1023. https://doi.org/10.1063/1.456153.

(91) Krauter, C. M.; Carter, E. A. EmbeddingIntegralGenerator. https://github.com/EACcodes/EmbeddingIntegralGenerator.

(92) Sun, Q.; Zhang, X.; Banerjee, S.; Bao, P.; Barbry, M.; Blunt, N. S.; Bogdanov, N. A.; Booth, G. H.; Chen, J.; Cui, Z.-H.; Eriksen, J. J.; Gao, Y.; Guo, S.; Hermann, J.; Hermes, M. R.; Koh, K.; Koval, P.; Lehtola, S.; Li, Z.; Liu, J.; Mardirossian, N.; McClain, J. D.; Motta, M.; Mussard, B.; Pham, H. Q.; Pulkin, A.; Purwanto, W.; Robinson, P. J.; Ronca, E.; Sayfutyarova, E. R.; Scheurer, M.; Schurkus, H. F.; Smith, J. E. T.; Sun, C.; Sun, S.-N.;

Upadhyay, S.; Wagner, L. K.; Wang, X.; White, A.; Whitfield, J. D.; Williamson, M. J.; Wouters, S.; Yang, J.; Yu, J. M.; Zhu, T.; Berkelbach, T. C.; Sharma, S.; Sokolov, A. Yu.; Chan, G. K.-L. Recent Developments in the PySCF Program Package. *J. Chem. Phys.* **2020**, *153* (2), 024109. https://doi.org/10.1063/5.0006074.

(93) Ye, H.-Z. PySCF. https://github.com/pyscf/pyscf-forge.

(94) Zeng, J.; Zhang, D.; Peng, A.; Zhang, X.; He, S.; Wang, Y.; Liu, X.; Bi, H.; Li, Y.; Cai, C.; Zhang, C.; Du, Y.; Zhu, J.-X.; Mo, P.; Huang, Z.; Zeng, Q.; Shi, S.; Qin, X.; Yu, Z.; Luo, C.; Ding, Y.; Liu, Y.-P.; Shi, R.; Wang, Z.; Bore, S. L.; Chang, J.; Deng, Z.; Ding, Z.; Han, S.; Jiang, W.; Ke, G.; Liu, Z.; Lu, D.; Muraoka, K.; Oliaei, H.; Singh, A. K.; Que, H.; Xu, W.; Xu, Z.; Zhuang, Y.-B.; Dai, J.; Giese, T. J.; Jia, W.; Xu, B.; York, D. M.; Zhang, L.; Wang, H. DeePMD-Kit v3: A Multiple-Backend Framework for Machine Learning Potentials. *J. Chem. Theory Comput.* **2025**, *21* (9), 4375–4385. https://doi.org/10.1021/acs.jctc.5c00340.

(95) Invernizzi, M.; Parrinello, M. Exploration vs Convergence Speed in Adaptive-Bias Enhanced Sampling. *J. Chem. Theory Comput.* **2022**, *18* (6), 3988–3996. https://doi.org/10.1021/acs.jctc.2c00152.

(96) Thompson, A. P.; Aktulga, H. M.; Berger, R.; Bolintineanu, D. S.; Brown, W. M.; Crozier, P. S.; in 't Veld, P. J.; Kohlmeyer, A.; Moore, S. G.; Nguyen, T. D.; Shan, R.; Stevens, M. J.; Tranchida, J.; Trott, C.; Plimpton, S. J. LAMMPS - a Flexible Simulation Tool for Particle-Based Materials Modeling at the Atomic, Meso, and Continuum Scales. *Comput. Phys. Commun.* **2022**, *271*, 108171. https://doi.org/10.1016/j.cpc.2021.108171.

(97) Tribello, G. A.; Bonomi, M.; Branduardi, D.; Camilloni, C.; Bussi, G. PLUMED 2: New Feathers for an Old Bird. *Comput. Phys. Commun.* **2014**, *185* (2), 604–613. https://doi.org/10.1016/j.cpc.2013.09.018.

(98) Carter, E. A.; Ciccotti, G.; Hynes, J. T.; Kapral, R. Constrained Reaction Coordinate Dynamics for the Simulation of Rare Events. *Chem. Phys. Lett.* **1989**, *156* (5), 472–477. https://doi.org/10.1016/S0009-2614(89)87314-2.

(99) DEN OTTER, W. K.; BRIELS, W. J. Free Energy from Molecular Dynamics with Multiple Constraints. *Mol. Phys.* **2000**, *98* (12), 773–781. https://doi.org/10.1080/00268970009483348.

(100) O'Neill, N.; Shi, B. X.; Witt, W. C.; Armstrong, B. I.; Baldwin, W. J.; Raiteri, P.; Schran, C.; Michaelides, A.; Gale, J. D. From Accurate Quantum Chemistry to Converged Thermodynamics for Ion Pairing in Solution. arXiv March 6, 2026. https://doi.org/10.48550/arXiv.2603.06800.